\documentclass[11pt]{article}
\usepackage{jheppub}

\usepackage{epsfig}
\usepackage{graphicx}
\def\be{\begin{equation}}
\def\ee{\end{equation}}
\def\bea{\begin{array}}
\def\eea{\end{array}}
\def\beqa{\begin{eqnarray}}
\def\eeqa{\end{eqnarray}}
\def\beqas{\begin{eqnarray*}}
\def\eeqas{\end{eqnarray*}}

\def\bp{\begin{picture}}
\def\ep{\end{picture}}
\def\bc{\begin{center}}
\def\ec{\end{center}}
\def\bfig{\begin{figure}}
\def\efig{\end{figure}}

\def\bit{\begin{itemize}}
\def\eit{\end{itemize}}
\def\nn{\nonumber}
\def\f{\frac}

\def\[{\left[}
\def\]{\right]}
\def\({\left(}
\def\){\right)}

\def\..{\left.}
\def\.{\right.}
\def\tl{\tilde}
\def\ra{\rightarrow}
\def\la{\leftarrow}

\def\NPB#1,{{ Nucl.\ Phys.\ B }{\bf #1},}
\def\PLB#1,{{ Phys.\ Lett.\ B }{\bf #1},}
\def\EPJC#1,{{ Eur.\ Phys.\ Jour.\ C }{\bf #1},}
\def\PRD#1,{{ Phys.\ Rev.\ D }{\bf #1},}
\def\PRL#1,{{ Phys.\ Rev.\ Lett.\ }{\bf #1},}
\def\MPLA#1,{{Mod.\ Phys.\ Lett.\ A }{\bf #1},}

\def\da{\dagger}

\def\la{\lambda}

\def\al{\alpha}

\def\ka{\kappa}

\def\ep{\epsilon}

\def\pa{\partial}
\def\pr{\prime}

\begin{document}

\title{ Analytical Soft SUSY Spectrum in Mirage-Type Mediation Scenarios}

\author[a,b]{Fei Wang,} \emailAdd{feiwang@zzu.edu.cn}

\affiliation[a]{Department of Physics and Engineering, Zhengzhou University, Zhengzhou 450000, P. R. China}

\affiliation[b]{State Key Laboratory of Theoretical Physics, Institute of Theoretical Physics,
Chinese Academy of Sciences, Beijing 100190, P. R. China}

\abstract{  We derive explicitly the soft SUSY breaking parameters at arbitrary low energy scale in the (deflected) mirage type mediation scenarios
with possible gauge or Yukawa mediation contributions. Based on the Wilsonian effective action after integrating out the messengers,
we obtain analytically the boundary value (at the GUT scale) dependencies of the effective wavefunctions and gauge kinetic terms.
Note that the messenger scale dependencies of the effective wavefunctions and gauge kinetic terms had already been discussed in GMSB.
The RGE boundary value dependencies, which is a special feature in (deflected) mirage type mediation,
 is the key new ingredients in this study.
The appearance of $'mirage'$ unification scale in mirage mediation is proved rigorously with our analytical results.
We also discuss briefly the new features in deflected mirage mediation scenario in the case the deflection comes purely from the Kahler potential
and the case with messenger-matter interactions.
}

\maketitle
\section{Introduction}

  After the discovery of the 125 GeV Higgs boson in 2012 at the CERN LHC\cite{ATLAS:higgs,CMS:higgs}, the long missing particle content of the Standard Model(SM)
has finally been verified. In spite of the impressive triumph of SM, many physicists still believe that new physics may be revealed at LHC.
Among the many new physics models that can solve the fine-tuning problem, the most elegant and compelling resolution is low energy supersymmetry.
Augmented with weak scale soft SUSY breaking terms, the quadratic cutoff dependence is absent, leaving only relatively mild but intertwined logarthmic sensitivity
to high scale physics.  As such soft SUSY breaking spectrum is determined by the SUSY breaking mechanism, it is interesting to survey the the phenomenology
related to supersymmetry breaking mechanism.

 In Type IIB string theory compactified on a Calabi-Yau (CY) orientifold, the presence of NS and RR 3-form background fluxes can fix the dilaton and 
the complex structure
moduli, leaving only the Kahler moduli in the  Wilsonian effective supergravity action
after integrating out the superheavy complex structure moduli and dilaton.
The remaining Kahler moduli fields could be stabilized by non-perturbative effects, such as instanton or gaugino condensation.
 In order to generates SUSY breaking in the observable sector and obtain a very tiny positive cosmological constant, Kachru-Kallosh-Linde-
Trivedi (KKLT)\cite{KKLT} propose to add an anti-D3 brane  at the tip of the Klebanov-Strassler throat (or adding F-term, D-term SUSY breaking contributions\cite{F-D-lifting})
to explicitly break SUSY and lift the AdS universe to obtain a dS one.
 In addition to the anomaly mediation contributions, SUSY breaking effects from the light Kahler moduli fields could also be mediated
to the visible sector and result in a mixed modulus-anomaly mediation SUSY breaking
scenario \cite{mirage:hep-th:0411066,mirage:hep-th:0503216}. It is interesting to note that the involved
modulus mediated SUSY breaking contributions can be comparable to that of the anomaly mediation \cite{AMSB}.
With certain assumptions on the Yukawa couplings and the modular weights, the SUSY breaking contributions
from the renormalization group running and anomaly mediation could cancel each other at a $'mirage'$ unification
scale, leading to a compressed low energy SUSY breaking spectrum \cite{mirage:hep-ph:0504037}.
Such a mixed modulus-anomaly mediation SUSY breaking mechanism is dubbed as $'{\rm mirage~ mediation}'$.

Anomaly mediation contribution is a crucial ingredient of such a mixed modulus-anomaly mediation.
It is well known that the pure anomaly mediation is bothered by the tachyonic slepton problem \cite{tachyonslepton}.
One of its non-trivial extensions with messenger sectors, namely the deflected anomaly mediated SUSY breaking (AMSB),
can elegantly solve such a tachyonic slepton problem through the deflection of the renormalization group equation (RGE) trajectory \cite{dAMSB,okada,fei}.
Such a messenger sector can also be present in the mirage mediation so that
additional gauge contributions by the messengers\cite{deflectmirage} can deflect the RGE trajectory
and change the low energy soft SUSY predictions. Additional deflection in mirage mediation can be advantageous in phenomenological aspect. 
For example, apparent gaugino mass unification at TeV scale could still be realized with the simplest $'no~scale'$ Kahler potential, which, 
in ordinary mirage mediation, can only be possible with the not UV-preferable $\al=2$ case. Relevant discussions on mirage-type mediation scenarios can be seen, 
for example, in\cite{dmirage1,dmirage2,mirageNMSSM1,mirageNMSSM2}.

In mirage type mediation scenarios, analytical expressions for the soft SUSY breaking parameters are no not given at the messenger scale $M$ (or scale below $M$), 
but given at the GUT scale instead.  One needs to numerically evolve the spectrum with GUT scale input to obtain the low energy SUSY spectrum. 
This procedure obscures the appearance of $'mirage'$ unification scale from the input. In mirage mediation scenarios with deflection from Kahler potential, 
analytical results of mirage mediation are necessary to predict the low energy SUSY spectrum. So it is preferable to give the analytical expressions for 
the soft SUSY breaking parameters in mirage type mediation scenarios at arbitrary low energy scale.  Besides, possible new Yukawa-type interactions involving 
the messengers may give additional Yukawa mediation contributions to the low energy soft SUSY spectrum (See \cite{fei:mirageNMSSM} for example). Such a generalization 
of deflected mirage mediation scenario shows new features in phenomenological studies. The inclusion of Yukawa mediation contributions at (or below) 
the messenger scale $M$ are non-trivial and again prefer analytical expressions near the messenger scale.

This paper is organized as follows. We briefly review the mirage type mediation scenarios in Sec.\ref{sec-2}. A general discussion on the analytical expressions 
for the soft SUSY parameters in the generalized deflected mirage mediation is given in Sec.\ref{sec-3}. 
We discuss some applications of our analytical results in Sec.\ref{application}, including the proof of the $'mirage'$ unification scale in mirage mediation with 
our analytical results and the discussions on deflection from Kahler potential. Sec.\ref{conclusion} contains our conclusions.
\section{\label{sec-2}Brief Review of the Mirage Type Mediation Scenarios}

 Inspired by string-motivated KKLT approach to moduli stabilization within Type IIB string theory,  mirage mediation supersymmetry breaking is proposed, 
in which the modulus mediated supersymmetry breaking terms are suppressed by numerically a loop factor so that the anomaly
mediated terms can be competitive. 

After fixing and integrating out the dilaton and the complex structure moduli,  the four-dimensional Wilsonian effective supergravity action (defined at the 
boundary scale $\Lambda$) in terms of compensator field and a single Kahler modulus parameterizing the overall size of the
compact space\cite{mirage:hep-ph:0504037} is given as

\small
\begin{eqnarray}
e^{-1}{\cal L}=\int d^4\theta \[\phi^\da\phi \(-3 e^{-K/3}\)-(\phi^\da\phi)^2\bar{\theta}^2\theta^2 {\cal P}_{lift}\]
+\int d^2\theta \phi^3 W+ \int d^2\theta \f{f_i}{4} W_i^a W_i^a
\label{action}
\end{eqnarray}
\normalsize
with a holomorphic gauge kinetic term
\beqa
f_i=\f{1}{g_i^2}+i\f{\theta}{8\pi}.
\eeqa
The Kahler potential takes the form
\beqa
K&=&-3\ln(T+T^\da)+Z_X(T^\da,T) X^\da X+Z_\Phi(T^\da,T)\Phi^\da\Phi\nn\\
 & & +\sum\limits_{i}Z_{P_i,\bar{P}_i}(T^\da,T)\[ P_i^\da P_i+\bar{P}_i^\da \bar{P}_i\]~,
\eeqa
with the $'no-scale'$ kinetic term for the Kahler modulus $T$. The gauge kinetic term $f_i$, the messenger superfields $P_i$, the MSSM superfields $\Phi$ and 
the pseudo-moduli superfields are all assumed to depend non-trivially on the Kahler moduli $T$ as
\beqa
Z_X(T^\da,T)&=&\f{1}{(T^\da + T)^{n_X}}~,~~Z_\Phi(T^\da,T)=\f{1}{(T^\da + T)^{n_\Phi}}~,~\nn\\
f_i(T)&=& T^{l_i}~,~~~~~~~~~~~~~Z_{P_i,\bar{P}_i}(T^\da,T)=\f{1}{(T^\da + T)^{n_P}}~.
\eeqa
Choices of $n_X,n_\Phi,n_P,l_i$ depend on the location of the fields on the D3/D7 branes. Besides, universal $l_i=1$ are adopted in our scenario to keep gauge 
coupling unification, so the gauge fields should reside on the D7 brane.

The superpotential takes the most general form involving the KKLT setup\cite{KKLT}, the messenger sectors $W_M$ and visible sector $W_{\overline{MSSM}}$
\beqa
W=\(\omega_0-A e^{-aT}\)+W_{M}+W_{\overline{MSSM}}~,
\eeqa
where the first term is generated from the fluxes and the second term from non-perturbative effects,
such as gaugino condensation or D3-instanton. Within $W_M$, interactions between messengers and MSSM fields can 
possibly arise which will be discussed subsequently.
The modulus $T$, which is not fixed by the background flux, can be stabilized by non-perturbative gaugino
condensation with its VEV satisfying
\beqa
a~\Re{\langle T \rangle}\approx\ln \(\f{A}{\omega_0}\)\approx \ln\(\f{M_{Pl}}{m_{3/2}}\)\approx 4\pi^2~
\eeqa
up to ${\cal O}(\ln[{M_{Pl}}/{m_{3/2}}]^{-1})$. Boundary value of the soft SUSY breaking parameters at the GUT scale can be seen in \cite{mirage:hep-ph:0504037}.
\section{\label{sec-3}Analytical Expressions of Soft SUSY Breaking Parameters}
Mirage mediation can be seen as a typical mixed modulus-anomaly mediation SUSY breaking mechanism with each contribution of similar size. Adding a messenger 
sector will add additional gauge mediation contributions. Besides, upon the messenger thresholds, new Yukawa interactions involving the messengers could arise. 
Such interactions may cause new contributions to trilinear couplings and sfermion masses (As an example, see our previous work \cite{fei:mirageNMSSM}). Additional deflection with Yukawa mediation can be advantageous in several aspects.
\bit
\item  The value of trilinear coupling $|A_t|$ can be increased by additional contributions involving the new Yukawa interactions. 
Larger value of $A_t$ is always welcome in MSSM and NMSSM not only to accommodate the 125 GeV Higgs but also to reduce\cite{rnaturalsusy} 
the EW fine tuning\cite{EWFT} involved.
\item  As noted in \cite{fei:mirageNMSSM,NMSSM:AMSB,fei:NMSSM-damsb}, pure gauge mediation contributions are not viable to generate either trilinear couplings $A_\ka,A_\lambda$ or 
soft scalar masses $m_S^2$ for singlet superfields $S$ which are crucial to solve the mu-problem of NMSSM. Deflection with Yukawa interactions will 
readily solve such difficulty.
\eit

To take into account such Yukawa mediation contributions in soft SUSY breaking parameters, it is better to derive the most general results involving the deflection. 
There are two approaches to obtain the low energy SUSY spectrum in the (deflected) mirage type mediation scenario:
 \bit
\item  In the first approach, the mixed modulus-anomaly mediation soft SUSY spectrum is given
      by their boundary values at the GUT scale\cite{mirage:hep-ph:0504037}.
      Such a spectrum will receive additional contributions towards its RGE running to low energy scale,
      especially the threshold corrections related to the appearance of messengers\cite{0901.0052,1001.5261}.
      The soft SUSY breaking parameters are obtained by combing numerical RGE evolutions with threshold corrections.
      In \cite{0901.0052}, following this approach, some analytical expressions of the soft SUSY spectrum, for example the gaugino masses,
     are given. General expressions of the soft scalar masses and trilinear couplings are not given explicitly except for some simplified cases.

\item In the second approach which we will adopt, the soft SUSY spectrum at low energy scale
is derived directly from the low energy effective action.
We know that the SUGRA description in eq.(\ref{action}) can be seen as a Wilsonian effective action after
integrating out the complex structure moduli and dilaton field.
After the pseudo-modulus acquires a VEV and determines the messenger threshold, the messenger sector
can be integrated out to obtain a low energy effective action below the messenger threshold.
So we anticipate the Kahler metric $Z_\Phi$ and gauge kinetic $f_i$ will depend non-trivially
on the messenger threshold $M_{mess}^2/\phi^\da\phi$ and $M_{mess}/\phi$, respectively.
The resulting soft SUSY spectrum below the messenger threshold can be derived from the wavefunction
renormalization approach \cite{GMSB:wavefunction}. The main difficulty here is to find the boundary value dependencies
of the wavefunction and gauge kinetic term.

In this approach, the most general expressions for soft SUSY breaking parameters in deflected modulus-anomaly (mirage) mediation SUSY breaking mechanism 
are derived below. Ordinary mirage mediation results can be obtained by setting the deflection parameter $' d '$ to zero.

\bit
\item The gaugino masses are given by
\beqa
\label{gaugino}
M_i&=& -g_i^2\(\f{F_T}{2}\f{\pa}{\pa T} -\f{F_\phi}{2}\f{\pa}{\pa \ln\mu}+\f{d F_\phi}{2}\f{\pa}{\pa \ln |X|}\) f_a(T,\f{\mu}{\phi},\sqrt{\f{X^\da X}{\phi^\da\phi}})~,
\eeqa

\item The trilinear terms are given by
\beqa
\label{trilinear}
&&A_{Y_{abc}}\equiv A_{abc}/y_{abc}~\\
&=&\f{1}{2}\sum\limits_{i=a,b,c}\({F^T}\f{\pa}{\pa T}-{F_\phi}\f{\pa}{\pa\ln\mu}+{d F_\phi}\f{\pa}{\pa\ln |X|}\) \ln \[e^{-K_0/3}Z_i(\mu,X,T)\]~.\nn
\eeqa

\item The soft sfermion masses are given by
\beqa
\label{scalar}
-m^2_{soft}(\mu)&=&\left|\f{F_T}{2}\f{\pa}{\pa T}-\f{F_\phi}{2}\f{\pa}{\pa\ln\mu}+\f{d}{2} F_\phi\f{\pa}{\pa\ln |X|}\right|^2 \ln \[e^{-K_0/3}Z_i(\mu,X,T)\]~\\
&=&\(\f{|F_T|^2}{4}\f{\pa^2}{\pa T\pa T^*}+\f{F_\phi^2}{4}\f{\pa^2}{\pa (\ln\mu)^2}+\f{d^2F^2_\phi}{4}\f{\pa^2}{\pa (\ln |X|)^2}-\f{F_TF_\phi}{2}\f{\pa^2}{\pa T\pa\ln\mu}\right.~\nn\\
&&~~+\left.\f{dF_TF_\phi}{2}\f{\pa^2}{\pa T\pa\ln|X|}-\f{d F^2_\phi}{2}\f{\pa^2}{\pa\ln|X|\pa\ln\mu}\) \ln \[e^{-K_0/3}Z_i(\mu,X,T)\],\nn
\eeqa

\eit

\eit

From the previous general expressions, we can deduce the concrete analytical results for soft SUSY parameters.
In our notation, we define the modulus mediation part 
\beqa
M_0\equiv\f{F_T}{T+T^*}~,~~~q_{Y_{ijk}}\equiv 3-(n_i+n_j+n_k)~.
\eeqa
The gauge and Yukawa couplings are used in the form
\beqa
\al_i=\f{g_i^2}{4\pi}~,~~~\al_{\la_{ijk}}=\f{\la^2_{ijk}}{4\pi}~.
\eeqa
\subsection{Gaugino Mass}
The gaugino mass below the messenger scale can be obtained from Eqn.(\ref{gaugino}).
 At the GUT (compactification scale) $M_G$, the gauge coupling unification requires
\beqa
T^{l_a}= \f{1}{g^2(GUT)}~,
\eeqa
The gauge coupling at scale $\mu$ just below the messenger threshold $M$ is given as
\beqa
\f{1}{g_i^2(\mu)}&=&\f{1}{g_i^2(GUT)}+\f{b_i+\Delta b_i}{8\pi^2}\ln\f{M_G}{|X|}+\f{b_i }{8\pi^2}\ln\f{|X|}{\mu}~,\nn\\
&=& T^{l_a}+\f{b_i+\Delta b_i}{8\pi^2}\ln\f{M_G}{M}+\f{b_i}{8\pi^2}\ln\f{M}{\mu}~.
\eeqa
The derivatives are given as
\beqa
\f{\pa }{\pa \ln\mu}\(\f{1}{g_i^2(\mu)}\)&=&-\f{b_i}{8\pi^2}~,~~~~~~
\f{\pa }{\pa \ln M}\(\f{1}{g_i^2(\mu)}\)=-\f{\Delta b_i}{8\pi^2}~,
\eeqa
and
\beqa
&&\f{\pa }{\pa T}\(\f{1}{g_a^2(\mu)}\)=l_a T^{l_a-1}~
\Longrightarrow -2\f{1}{g_a^3}\f{\pa g_a(\mu)}{\pa T}=l_a T^{l_a-1}~,
\eeqa
So we can obtain the analytical results for gaugino mass
\beqa
M_i(\mu)=g_i^2(\mu)\[l_a\f{F_T}{2T}\f{1}{g_a^2(GUT)}+\f{F_\phi}{2}\f{b_i}{8\pi^2}-\f{d}{2} F_\phi\f{\Delta b_i}{8\pi^2}\]~.
\eeqa
with $\Delta b_i\equiv b_i^\pr-b_i$ and $b_i^\pr,~b_i$  the gauge beta function upon and below the messenger thresholds, respectively.
This results can coincide with the gaugino masses predicted from RGE running with threshold corrections at the messenger scale.
Following the approach in \cite{deflectmirage}, the gaugino mass at the scale $\mu$ slightly below the messenger scale $M$ will receive additional gauge mediation contributions
\beqa
M_i(\mu\lesssim  M)&=&\f{g_i^2( M)}{g_i^2(GUT)}M_i(GUT)-F_\phi \f{g_i^2( M)}{16\pi^2} (d+1) \Delta b_i~,\nn\\
&=&g_i^2( M)\[l_a\f{F_T}{2T}\f{1}{g_a(GUT)}+\f{F_\phi}{2}\f{b_i+\Delta b_i}{8\pi^2}\]-F_\phi \f{g_i^2( M)}{16\pi^2} (d+1) \Delta b_i,
\eeqa
with
\beqa
M_i(GUT)=g_i^2(GUT)\[l_a\f{F_T}{2T}\f{1}{g_a^2(GUT)}+\f{F_\phi}{2}\f{b_i+\Delta b_i}{8\pi^2}\].
\eeqa

Then we can obtain the gaugino mass at scale $\mu< M$ from one-loop RGE
\beqa
M_i(\mu)&=&\f{g_i^2(\mu)}{g_i^2( M)} M_i(\mu\lesssim\ln M)~,\nn\\
&=&g_i^2(\mu)\[l_a\f{F_T}{2T}\f{1}{g_a^2(GUT)}+\f{F_\phi}{2}\f{b_i}{8\pi^2}\]-F_\phi \f{g_i^2(\mu)}{16\pi^2} d \Delta b_i~,
\eeqa
So we can see that the two results agree with each other.
\subsection{Trilinear Terms}
From the form of wavefunction
\beqa
Z_i(\mu)=Z_i(\Lambda)\prod\limits_{l=y_t,y_b,y_\tau} \(\f{y_l(\mu)}{y_l(\Lambda)}\)^{A_l}\prod\limits_{k=1,2,3}\(\f{g_k(\mu)}{g_k(\Lambda)}\)^{B_k}
\eeqa
we can obtain the trilinear terms for scales below the messenger $M$ from Eqn.(\ref{trilinear}). The main challenge is the calculation of $\pa Z_i/\pa T$.

Before we derive the final results involving all $y_t,y_b,y_\tau$  and $g_3,g_2,g_1$, we will study first the simplest case in which only the top Yukawa 
$\al_t\equiv y_t^2/4\pi$ and $\alpha_s\equiv g_3^2/4\pi$ are kept in the anomalous dimension. The RGE equation for $\alpha_t$ and $\al_s$ takes the form
\beqa
\f{d}{dt}\ln \alpha_t&=&\f{1}{\pi}\(3\alpha_t-\f{8}{3}\al_s\)~,~~~~~~~~~~~~~
\f{d}{dt}\ln \al_s=-\f{1}{2\pi} b_3\al_s~,
\eeqa
Note the definition $b_3$ differs by a minus sign.
Define $A=\ln\(\alpha_t \al_s^{-\f{16}{3b_3}}\)$, the equation can be written as
\beqa
\f{d}{dt} e^{-A}=-\f{3}{\pi}\al_s^{\f{16}{3b_3}}~,
\eeqa
So we can exactly solve the differential equation to get
\beqa
\[\f{\alpha_t(\mu)}{\al_t(\Lambda)}\( \f{\al_s(\mu)}{\al_s(\Lambda)}\)^{-\f{16}{3b_3}}\]^{-1}
=1-\f{3\al_t(\Lambda)}{\pi}\f{2\pi}{\f{16}{3}-b_3}\[\al_s(\Lambda)^{-1}-\(\f{\al_s(\mu)}{\al_s(\Lambda)}\)^{\f{16}{3b_3}}\al^{-1}_s(\mu)\].
\eeqa
Expanding the expressions and neglect high order terms, we finally have
\beqa
\f{\pa}{\pa T} \[\ln {\alpha_t(\mu)}-\ln{\al_t(\Lambda)}\]
&\approx&\f{\pa}{\pa T}\[-\f{8}{3\pi}\al_s(\mu)+\f{3}{\pi}\al_t(\mu)\]\ln\(\f{\Lambda}{\mu}\).
\eeqa
after calculations. It can be observed that the expression within the square bracket is just the beta function of top Yukawa coupling.

Now we will calculate $\pa Z_i/\pa T$ with all $y_t,y_b,y_\tau$ and $g_3,g_2,g_1$ taking into account in the expression.
\bit

\item  Deduction of $\pa Z_i/\pa T$ without messenger deflections:

From the form of wavefunction
\beqa
Z_i(\mu)=Z_i(M_G)\prod\limits_{l=y_t,y_b,y_\tau} \(\f{y_l(\mu)}{y_l(M_G)}\)^{A_l}\prod\limits_{k=1,2,3}\(\f{g_k(\mu)}{g_k(M_G)}\)^{B_k}
\eeqa
and renormalizatoin $Z=Z_0(1+\delta Z)$, we have
\beqa
\f{\pa \ln e^{-K_0/3} Z_i}{\pa T}&=&\f{\pa}{\pa T}\ln e^{-K_0/3}Z_i(M_G)+\f{\pa }{\pa T} \delta Z_i~,\nn\\
&=&\f{1-n_i}{T} +\sum\limits_{m=1,2}\[\sum\limits_{a}\f{\pa g_{a;m}}{\pa T}\f{\pa \delta Z_i}{\pa g_{a;m}}
+\sum\limits_{a,b,c} \f{\pa \ln y_{abc;m}}{\pa T}\f{\pa \delta Z_i}{\pa \ln y_{abc;m}}\]~,\nn
\eeqa
with $m=1,2$ corresponding to the value at the scale $\mu$ and the $GUT$ scale, respectively.

The derivative with respect to $g_m$ gives
\beqa
\f{\pa \ln e^{-K_0/3} Z_i}{\pa g_m(\mu)}=\f{B_m}{g_m(\mu)}~,~~~~~\f{\pa \ln e^{-K_0/3} Z_i}{\pa g_m(M_G)}=-\f{B_m}{g_m(M_G)}~,
\eeqa
and
\beqa
\label{GT}
\f{\pa g_i(\mu)}{\pa T}=-\f{l_a T^{l_a-1}}{2}g^3_i(\mu)~,~~~~~\f{\pa g_i(M_G)}{\pa T}=-\f{l_a T^{l_a-1}}{2}g^3_i(M_G)~.
\eeqa
The derivative with respect to $y_l$ gives
\beqa
\f{\pa \ln e^{-K_0/3} Z_i}{\pa y_l(\mu)}=\f{A_l}{y_l(\mu)}~,~~~\f{\pa \ln e^{-K_0/3} Z_i}{\pa y_l(M_G)}=-\f{A_l}{y_l(M_G)}~,
\eeqa
and
\beqa
\label{YT}
\f{\pa y_l(M_G)}{\pa T}&=&-\f{y_l(M_G)}{2}\[\f{3-a_{ijk}}{T}\],~~~~~
\f{\pa \ln \al_{Y_{abc}}(\mu)}{\pa T}=-\[\f{3-a_{ijk}}{T}\]~~.
\eeqa

From the beta function of the Yukawa couplings, we have
\beqa
\f{\pa\ln \al_{Y_{abc}}(\mu)}{\pa T}&=&\f{\pa\ln \al_{Y_{abc}}(M_G)}{\pa T}-\f{\pa }{\pa T}\int\limits_{\mu}^{M_G}\( \f{d}{d\ln\mu^\pr} \ln \al_{Y_{abc}}\)~ d\ln\mu^\pr~,\nn\\
&=&-\f{3-a_{abc}}{T}-\f{1}{2\pi } \int\limits_{\mu}^{M_G} d\ln\mu^\pr \(\sum\limits_{Y_{lmn}}c_{lmn}\f{\pa}{\pa T}\al_{Y_{lmn}}(\mu^\pr)+\sum\limits_{m}d_m\f{\pa}{\pa T}\al_m(\mu^\pr)\)~,\nn\\
&\approx&-\f{3-a_{abc}}{T}+\f{1}{2\pi }\[ \sum\limits_{Y_{lmn}}c_{lmn}\f{3-a_{lmn}}{T}\al_{Y_{lmn}}(\mu)+\sum\limits_{m}d_m\f{l_a}{T}\f{\al_m^2(\mu)}{\al_m(M_G)}\]\ln\(\f{{M_G}}{\mu}\)~,\nn
\eeqa
with $a_{abc}=n_a+n_b+n_c$~.

So the derivative with respect to $T$ is given by
\beqa
\f{\pa }{\pa T} \delta Z_i&=&\sum\limits_{m=1,2}\[\sum\limits_{a}\f{\pa  \ln \al_{a;m}}{\pa T}\f{\pa \delta Z_i}{\pa \ln \al_{a;m}}
+\sum\limits_{Y_{abc}} \f{\pa \ln \al_{Y_{abc;m}}}{\pa T}\f{\pa \delta Z_i}{\pa \ln \al_{Y_{abc;m}}}\]~,\nn\\
&=&\sum\limits_{a}\f{B_a}{2}\[\f{\pa}{\pa T} \ln\(\f{\al_a(\mu)}{\al_a(\Lambda)}\)\]+\sum\limits_{Y_{abc}}\f{A_{Y_{abc}}}{2}\[\f{\pa}{\pa T} \ln\(\f{\al_{Y_{abc}}(\mu)}{\al_{Y_{abc}}(\Lambda)}\)\]~,\nn\\
&\approx&\sum\limits_{a}\f{B_a}{2}\[\f{\pa}{\pa T}\(-\f{b_a}{2\pi}\al_a(\mu)\ln\(\f{\Lambda}{\mu}\)\) \]~\nn\\
&+&\sum\limits_{Y_{abc}}\f{A_{Y_{abc}}}{2}\f{1}{2\pi }\[ \sum\limits_{Y_{lmn}}c_{lmn}\f{3-a_{lmn}}{T}\al_{Y_{lmn}}(\mu)-\sum\limits_{m}d_m\f{\pa}{\pa T}\al_m(\mu)\]\ln\(\f{\Lambda}{\mu}\)~,\nn\\
\eeqa
We know from the expression of the wavefunction, the coefficients satisfy
\beqa
\sum\limits_{Y_{abc}}\f{A_{Y_{abc}}}{2}d_m+  b_m \f{B_m}{2}=-\f{\pa G_{Z_i}}{\pa \al_m},
\eeqa
for coefficients of $\al_m$. While the coefficients for Yukawa couplings $Y_{lmn}$ within $Z_i$ satisfy
\beqa
\sum\limits_{Y_{abc}}\f{A_{Y_{abc}}}{2}c_{lmn}=-\f{\pa G_{Z_i}}{\pa \al_{Y_{lmn}}},
\eeqa
So the final results reduces to
\beqa
\f{\pa}{\pa T} \ln e^{-K_0/3}Z_i
&\approx&-\f{1}{2\pi}\[ \f{d_{jk}^i}{2} \f{3-a_{Y_{ijk}}}{T}\al_{Y_{ijk}}(\mu)-2C_a(i)\f{l_a}{T}{\al_a(\mu)}\]\ln\(\f{GUT}{\mu}\)+\f{1-n_i}{T}~,\nn\\
&=&\f{1}{2\pi}\f{\pa}{\pa T}\[\f{d_{jk}^i}{2}\al_{Y_{ijk}}(\mu)-2C_a(i)\al_a(\mu)\]\ln\(\f{GUT}{\mu}\)+\f{1-n_i}{T}
\eeqa
with the expressions in the second square bracket being the anomalous dimension of $Z_i$.
\item   Deduction of $\pa Z_i/\pa T$ with messenger deflections:

From the form of wavefunction
\beqa
Z_i(\mu)&=&Z_i({M_G})\prod\limits_{l=y_t,y_b,y_\tau} \(\f{y_l(M)}{y_l({M_G})}\)^{A_l}\prod\limits_{k=1,2,3}\(\f{g_k(M)}{g_k({M_G})}\)^{B_k}\prod\limits_{k=y_U} \(\f{y_k(M)}{y_k({M_G})}\)^{C_k}\nn\\
&&~~~~~~~~\prod\limits_{l=y_t,y_b,y_\tau} \(\f{y_l(\mu)}{y_l(M)}\)^{A_l^\pr}\prod\limits_{k=1,2,3}\(\f{g_k(\mu)}{g_k(M)}\)^{B_k^\pr}~,
\eeqa
with $y_U$ the interactions involving the messengers which will be integrated below the messenger scale.

We have
\small
\beqa
\f{\pa \ln e^{-K_0/3} Z_i}{\pa T}
&=&\f{\pa}{\pa T}\ln e^{-K_0/3}Z_i^0+\f{\pa }{\pa T} \delta Z_i~,\nn\\
&=&\[\sum\limits_{g_a}\f{\pa \ln \(\f{g_{a}(\mu)}{g_a(M)}\)}{\pa T}\f{\pa \delta Z_i}{\pa \ln \(\f{g_{a}(\mu)}{g_a(M)}\)}
+\sum\limits_{y_{abc}} \f{\pa \ln\(\f{y_{abc}(\mu)}{y_{abc}(M)} \)}{\pa T}\f{\pa \delta Z_i}{\pa \ln\(\f{y_{abc}(\mu)}{y_{abc}(M)} \)}
\right.~,\nn\\
&+&\sum\limits_{g_a}\f{\pa \ln \(\f{g_{a}(M)}{g_a({M_G})}\)}{\pa T}\f{\pa \delta Z_i}{\pa \ln \(\f{g_{a}(M)}{g_a({M_G})}\)}
+\sum\limits_{y_{abc}} \f{\pa \ln\(\f{y_{abc}(M)}{y_{abc}({M_G})} \)}{\pa T}\f{\pa \delta Z_i}{\pa \ln\(\f{y_{abc}(M)}{y_{abc}({M_G})} \)},\nn\\
&+&\left.\sum\limits_{y_U} \f{\pa \ln\(\f{y_{U}(M)}{y_{U}({M_G})} \)}{\pa T}\f{\pa \delta Z_i}{\pa \ln\(\f{y_{U}(M)}{y_{U}({M_G})} \)}\right]+\f{1-n_i}{T}~,\nn\\
\eeqa
\normalsize
with $m=1,2$ corresponding to the value at the scale $\mu$ and the $GUT$ scale, respectively.

Using similar deductions for Yukawa couplings,  we can obtain
\small
\beqa
\f{\pa \ln e^{-K_0/3} Z_i}{\pa T}
&=&\f{1-n_i}{T}-\f{1}{4\pi}\sum\limits_{g_a} \( B_{a}  b^\pr_a \f{\pa \al_a(M)}{\pa T}\ln\(\f{M_G}{M}\)
+B_{a}^\pr b_a \f{\pa  \al_a(M) }{\pa T}\ln\(\f{M }{\mu}\)\)\nn\\
&+&\sum\limits_{y_{abc}\in y_t,y_b,y_\tau} A_{y_{abc}}\ln\(\f{{M_G}}{M}\) \f{1}{4\pi }\[ \sum\limits_{Y_{lmn}\in y_t,y_b,y_\tau}c_{lmn}\f{3-a_{lmn}}{T}\al_{Y_{lmn}}(M)\right. ~,\nn\\
&& \left.~~~~~~~~~~~~~~~~~~~~~~~~~~~+\sum\limits_{ {Y}_{lmn}\in y_U}\tl{c}_{lmn}\f{3-a_{U}}{T}\al_{Y_U}(M)-\sum\limits_{g_m}d_m\f{\pa}{\pa T}\al_m(M)\] ~.\nn\\
&+&\sum\limits_{y_{abc}\in y_t,y_b,y_\tau}A_{y_{abc}}^\pr\ln\(\f{\mu}{M}\)\f{1}{4\pi }\[ \sum\limits_{Y_{lmn}\in y_t,y_b,y_\tau}c_{lmn}\f{3-a_{lmn}}{T}\al_{Y_{lmn}}(M)-\sum\limits_{g_m}d_m\f{\pa}{\pa T}\al_m(M)\] ~.\nn\\
&+&\sum\limits_{y_U} C_{y_U} \ln\(\f{M}{{M_G}}\)\f{1}{4\pi }\[ \sum\limits_{\tl{Y}_{lmn}\in y_t,y_b,y_\tau}d_{lmn}\f{3-a_{lmn}}{T}\al_{Y_{lmn}}(M)-\sum\limits_{g_m}f_m\f{\pa}{\pa T}\al_m(M)~,\right.\nn\\
&& \left.~~~~~~~~~~~~~~~~~~~~~~~~~~~+\sum\limits_{\tl{Y}_{lmn}\in y_U}\tl{d}_{lmn}\f{3-a_{U}}{T}\al_{Y_U}(M)\]~.
\eeqa\normalsize
with the beta function for $y_t,y_b,y_\tau$ Yukawa couplings
\beqa
16\pi^2 \beta_{Y_{abc}}(\mu)&=&\left\{ \bea{c}\sum\limits_{Y_{lmn}\in y_t,y_b,y_\tau}c_{lmn}\al_{Y_{lmn}}
+\sum\limits_{Y_{lmn}\in y_U}\tl{c}_{lmn}\al_{Y_{U}}-\sum\limits_{g_m}d_m \al_m~,~~~~~ \mu \gtrsim M,\nn\\
 \sum\limits_{Y_{lmn}\in y_t,y_b,y_\tau}c_{lmn}\al_{Y_{lmn}}-\sum\limits_{g_m}d_m \al_m~,~~~~ \mu \lesssim M,\nn\\  \eea\right.
\eeqa
and the beta function for new messenger-matter $y_U$ Yukawa couplings
\beqa
16\pi^2 \beta_{Y_{U}}&=& \sum\limits_{\tl{Y}_{lmn}\in y_t,y_b,y_\tau}d_{lmn}\al_{\tl{Y}_{lmn}}
+\sum\limits_{\tl{Y}_{lmn}\in y_U}\tl{d}_{lmn}\al_{Y_{U}}-\sum\limits_{m}f_m \al_m~.
\eeqa

The coefficients satisfy
\beqa
&&\sum\limits_{Y_{abc}}\(A_{y_{abc}}-A_{y_{abc}}^\pr\) c_{lmn}+\sum\limits_{Y_U}C_{y_U}d_{lmn}=0~,~~~~~~({\rm for} ~y_t,y_b,y_\tau~ {\rm coefficients})\nn\\
 &&\sum\limits_{Y_{abc}} A_{y_{abc}}\tl{c}_{lmn} + \sum\limits_{Y_U} C_{y_U} \tl{d}_{lmn}=0~,~~~~({\rm for ~y_U~ coefficients})\nn\\
 && B_{m}  b^\pr_m+ \sum\limits_{Y_{abc}} A_{y_{abc}} d_m+\sum\limits_{Y_U} C_{Y_U} f_m=
 B^\pr_{m}  b_m+ \sum\limits_{Y_{abc}} A^\pr_{y_{abc}} d_m
\eeqa
and similarly for $g_m$, the sum then reduces to the previous case. So we have for $\mu<M$
\beqa
\label{DT}
&&\f{\pa}{\pa T} \ln \(e^{-K_0/3} Z_i(\mu)\)-\f{1-n_i}{T}~,\\
& \approx&-\f{1}{2\pi}\[ \f{1}{2}d_{jk}^i \f{3-a_{Y_{ijk}}}{T}\al_{Y_{ijk}}(\mu)-2C_a(i)\f{l_a}{T}\al_a(\mu)\]\ln\(\f{M_G}{\mu}\)~.\nn
\eeqa
Note that the expressions within the square bracket agree with the anomalous dimension of $Z_i^-$ below the messenger threshold $M$
\beqa
G_i^-\equiv \f{d Z_{i}^-}{d\ln\mu}\equiv-\f{1}{8\pi^2}\( \f{1}{2} d_{kl}^i\la^2_{ikl}-2c_r^ig_r^2\).
\eeqa
The $G_i^+$, which is  the anomalous dimension of $Z_i$ upon the messenger threshold $M$, do not appear in the final expressions.

\eit
The dependence of $Z_i$ on messenger scale $M$ can be derived following the techniques \cite{chacko,shih} developed in 
gauge mediated SUSY breaking (GMSB)\cite{GMSB} scenarios. From the expressions of the wavefunction, we can obtain
\beqa
\f{\pa}{\pa \ln M}\ln \[e^{-K_0/3}Z_i\]&=&\f{1}{4\pi}\sum\limits_{g_k}\[(B_k-B_k^\pr) (b_k+N_F) \al_k(M)+B_k^\pr N_F \al_k(\mu)\]\\
&+&\sum\limits_{Y_l}\[({A_l-A_l^\pr})G^+_{Y_l}(\ln M)+ {A_l^\pr}\f{\pa Y_l(\ln \mu,M)}{\pa \ln M}\]+
\sum\limits_{Y_U}\[{C_l} G^+_{Y_U}(\ln M)\]~.\nn
\eeqa
So the main challenge is to calculate $\pa \ln Y_a(\mu,\ln M)/\pa \ln M$.

From the beta functions for Yukawa couplings upon and below the messenger thresholds, the Yukawa couplings at scale $\mu<M$ is given as
\beqa
\ln Y_a(\mu,\ln M)=\ln Y_a({M_G})+\int\limits_{{M_G}}^{\ln M} G^+_{Y_a}(t^\pr) dt^\pr+\int\limits_{\ln M}^{\ln\mu} G^-_{Y_a}(t^\pr, \ln M) dt^\pr~,
\eeqa
with the Yukawa beta functions expressed as
\beqa
\beta_{Y_a}\equiv G_{Y_a} &\equiv&-\f{1}{2}\sum\limits_{i\in a} G^i\equiv\f{1}{4\pi}\(\f{1}{2}\tl{d}_{kl}^i\al_{\la_{ikl}}-2\tl{c}_r\al_r\)~,~\nn\\
G_i=\f{d \ln Z_{i}}{d\ln\mu}&\equiv&-\f{1}{2\pi}\(\f{1}{2}d_{kl}^i\al_{\la_{ikl}}-2c_r^i\al_r\).
\eeqa

We can derive the Yukawa couplings dependence on $'\ln M'$ at scale $\mu<M$
\beqa
\f{\pa}{\pa \ln M} \ln Y_a(\mu,\ln M)&=&\[ G^+_{Y_a}(\ln M)-G_{Y_a}^-(\ln M,\ln M)\]+\int\limits_{\ln M}^{\ln\mu} \f{\pa}{\pa \ln M}G^-_{Y_a}(t^\pr, \ln M) dt^\pr~,\nn\\
&\approx&\Delta G_{Y_a}(\ln M)-\f{1}{16\pi^2}\[\tl{d}_{kl}^i\la_{ikl}(\mu)\Delta G_{\la_{ikl}}-4\tl{c}_r \f{\Delta b_r}{16\pi^2} g_r^4(\mu)\]\ln\(\f{M}{\mu}\)~,\nn
\eeqa
In the case $\Delta G=0$ in which no additional Yukawa couplings involving the messengers are present, we have
\beqa
\f{\pa}{\pa \ln M}\ln Y_a(\mu,\ln M)
&\approx& \f{\tl{c}_r}{4\pi^2}  {\Delta b_r} \al_r^2(\mu)\ln\(\f{M}{\mu}\)~.
\eeqa
Note that at the messenger scale
\beqa
\f{\pa}{\pa \ln M}\ln Y_a(\ln M,\ln M)=\Delta G_a(\ln M).
\eeqa

The expressions takes a simple form at the scale $\mu$ slightly below the messenger scale $M$
\beqa
&& A_{Y_{abc}}(\mu\lesssim M)-\(3-a_{abc}\)\f{F_T}{T+T^*}\nn\\
&=&\sum\limits_{l=a,b,c}\left\{-\f{F_T}{T+T^*}\f{1}{2\pi}\[ \f{1}{2}d_{jk}^i (3-a_{Y_{ijk}})\al_{Y_{ijk}}(\mu)-2C_a(i){l_a}\al_a(\mu)\]\ln\(\f{GUT}{\mu}\)\right.\nn\\
 && \left.+d F_\phi\f{\Delta G_i}{2}-\f{F_\phi}{2} G_i^-\f{}{}\right\}~,\nn
\eeqa
with $\Delta G_i\equiv G_i^+-G_i^-$ [here $'G_i^+(G_i^-)'$ denotes respectively the anomalous dimension of $Z_i$ upon (below) the messenger threshold] 
the discontinuity of anomalous dimension across the messenger threshold.

\subsection{Soft Scalar Masses}

The soft scalar masses are given as
\beqa
\label{sfermion}
-m^2_{soft}&=&\left|\f{F_T}{2}\f{\pa}{\pa T}-\f{F_\phi}{2}\f{\pa}{\pa\ln\mu}+d F_\phi\f{\pa}{\pa\ln X}\right|^2 \ln \[e^{-K_0/3}Z_i(\mu,X,T)\]~,\nn\\
&=&\(\f{|F_T|^2}{4}\f{\pa^2}{\pa T\pa T^*}+\f{F_\phi^2}{4}\f{\pa^2}{\pa (\ln\mu)^2}+\f{d^2F^2_\phi}{4}\f{\pa}{\pa (\ln |X|)^2}-\f{F_TF_\phi}{2}\f{\pa^2}{\pa T\pa\ln\mu}\right.~\nn\\
&&~~+\left.\f{dF_TF_\phi}{2}\f{\pa^2}{\pa T\pa\ln|X|}-\f{d F^2_\phi}{2}\f{\pa^2}{\pa\ln|X|\pa\ln\mu}\) \ln \[e^{-K_0/3}Z_i(\mu,X,T)\],
\eeqa
The new ingredients are the second derivative of $Z_i$ with respect to $T$
\beqa
\label{DTT}
&&\f{\pa^2}{\pa T^2} \ln \[e^{-3K_0} Z_i\]\nn\\
&=&-\f{1}{2\pi}\f{\pa}{\pa T}\[\f{1}{2} d_{jk}^i \f{3-a_{Y_{ijk}}}{T}\al_{Y_{ijk}}(\mu)-2C_a(i)\f{l_a}{T}{\al_a(\mu)}\]\ln\(\f{GUT}{\mu}\)-\f{1-n_i}{T^2}~,\nn\\
&=&-\f{1}{2\pi}\[\f{1}{2}d_{jk}^i \f{3-a_{Y_{ijk}}}{T}\al_{Y_{ijk}}(\mu)\[-\f{3-a_{Y_{ijk}}}{T}+\f{1}{2\pi}\(\f{\tl{d}^p_{mn}}{2}\f{3-a_{Y_{ijk}}}{T}\al_{Y_{mnp}}-2c_r\f{l_a}{T}{\al_a}\)
\ln\(\f{GUT}{\mu}\)\]\right.\nn\\
 &-& \left. \f{1}{2}d_{jk}^i \f{(3-a_{Y_{ijk}})}{T^2}\al_{Y_{ijk}}(\mu)- 2C_a(i)\(-\f{l_a}{T^2}{\al_a(\mu)}-\f{l_a^2}{T^2}\f{\al_a^2(\mu)}{\al_a(GUT)}\)\]\ln\(\f{GUT}{\mu}\)-\f{1-n_i}{T^2}~.\nn\\
\eeqa
with the beta function of $Y_{ijk}$ given by
\beqa
\f{d \ln Y_{ijk}}{d\ln\mu}=\f{1}{16\pi^2}\[\f{\tl{d}^p_{mn}}{2}\al_{Y_{mnp}}-2c_r^i{\al_i}\].
\eeqa
and
\beqa
\f{\al_a(\mu)}{\al_a(GUT)}=1-\f{b_a}{2\pi}\al_a(\mu)\ln\(\f{GUT}{\mu}\).
\eeqa

The other terms within Eqn.(\ref{sfermion}) can be found in GMSB (not involving $\pa T$ ) or calculated directly using Eqn.(\ref{GT}) and Eqn.(\ref{YT}) 
(involving $\pa T$ ). We list the analytical results of deflected mirage mediation in Appendix B.

\section{\label{application} Applications Of The General Analytical Results}
\subsection{\label{sec-4}Analytical Results for Mirage Mediation}
Equipped with the previous deduction, we can readily reproduce the ordinary mirage mediation results by setting $d\ra 0$.
As the visible gauge fields originate from D7 branes and gauge coupling unification is always assumed, we adopt $l_a=1$.
The following definitions are used
\beqa
M_0\equiv\f{F_T}{2T}\equiv\f{F_\phi}{\al\ln\(\f{M_{Pl}}{m_{3/2}}\)}\approx \f{F_\phi}{4\pi^2\al}~.
\eeqa
with the parameter $\al$ defined as the ratio between the anomaly mediation and modulus mediation contributions
and the approximation  $\ln ({M_{Pl}}/{m_{3/2}}) \approx 4 \pi^2$.
We have
\bit
\item Gaugino mass:
\beqa
M_i(\mu)&=&l_a M_0\f{g_i^2(\mu)}{g_a^2(GUT)}+\f{F_\phi}{16\pi^2}b_i g_i^2(\mu)~,\nn\\
&=&l_a M_0 \[ 1-\f{b_i}{8\pi^2}g_i^2(\mu)\ln\f{GUT}{\mu}\]+\f{M_0}{4}\al b_i g_i^2(\mu)~.
\eeqa
So we can see that at the  scale $\mu_{Mi}$ which satisfies
\beqa
\f{1}{8\pi^2}\ln\(\f{M_{GUT}}{\mu_{Mi}}\)=\f{\al}{4}.
\eeqa
the gaugino masses unify at such $'mirage'$ unification scale
\beqa
\mu_{Mi}=M_{GUT} e^{-2\al \pi^2}\approx M_{GUT} \(\f{m_{3/2}}{M_{Pl}}\)^{\f{\al}{2}}.
\eeqa
\item Trilinear Term:
\beqa
&& A_{Y_{abc}}(\mu\lesssim M)\nn\\
&=&\sum\limits_{l=a,b,c}\left\{-M_0\f{1}{2\pi}\[ \f{1}{2}d_{jk}^i (3-a_{Y_{ijk}})\al_{Y_{ijk}}(\mu)-2C_a(i){l_a}\al_a(\mu)\]\ln\(\f{GUT}{\mu}\)\right.\nn\\
 && +\f{F_\phi}{4\pi}\[\f{1}{2}d_{jk}^i \al_{Y_{ijk}}(\mu)-2C_a(i)\al_a(\mu)\]+\(3-a_{abc}\)M_0~,\nn
\eeqa
In case the effect of Yukawa couplings are negligible or $a_{Y_{ijk}}=2$, the trilinear term also $"unify"$ at a mirage scale at which the last two terms 
cancel
\beqa
\f{1}{2\pi}\ln\(\f{M_{GUT}}{\mu_{Mi}}\)=\pi{\al}.
\eeqa
which is just the mirage scale for gaugino mass $"unification"$.

\item Soft Scalar Masses:
\small
\beqa
-m_i^2
&=&\f{M_0^2}{2\pi}\ln\(\f{M_{GUT}}{\mu}\)\left\{ \f{d_{jk}^i}{2}\(  q_{Y_{ijk}}^2+q_{Y_{ijk}}\)\al_{Y_{ijk}}(\mu)- 2C_a(i)\({l_a}+{l_a^2}\)\al_a\right.\nn\\
&+&\left.\f{1}{2\pi}\[\f{d_{jk}^i}{2}\al_{Y_{ijk}}(\mu)\(-\f{\tl{d}^p_{mn}}{2}q_{Y_{mnp}}\al_{Y_{mnp}}+2c_r l_a{\al_a}\)+2C_a(i) b_a \al^2_a\]
\ln\(\f{GUT}{\mu}\) \right\}\nn\\
 &+&\f{M_0 F_\phi}{2\pi}\[\f{d_{jk}^i}{2}\al_{Y_{ijk}}\(-q_{Y_{ikl}}
+\f{1}{2\pi}\[\f{\tl{d}^p_{mn}}{2} {q_{Y_{mnp}}}\al_{Y_{mnp}}-2c_r {l_r} {\al_r}\]\ln\(\f{M_{GUT}}{\mu}\)\) +2C_a(i) l_a \f{\al_a^2}{\al_a(GUT)}\]\nn\\
&-&\f{F_\phi^2}{8\pi}\[\f{d_{jk}^i}{2}\f{1}{2\pi}\(\f{\tl{d}^p_{mn}}{2}\al_{Y_{mnp}}-2c_r{\al_r}\) \al_{Y_{ijk}}- 2C_a(i)\f{b_a}{2\pi}\al_a^2\]-(1-n_i)M_0^2.
\eeqa
\normalsize
with $q_{Y_{ijk}}\equiv 3-(n_i+n_j+n_k)=3-a_{ijk}$.
Again, we can check that for $q_{Y_{ijk}}=1$ or negligible Yukawa couplings, the soft scalar masses apparent unify at $\mu_{Mi}$ defined above
\small
\beqa
&&-m_i^2+ (1-n_i)M_0^2\nn\\
&=& \pi\al M_0^2 \left\{ 2\f{d_{jk}^i}{2}\al_{Y_{ijk}}(\mu)- 4C_a(i)\al_a+
\f{1}{2\pi}\[\f{d_{jk}^i}{2}\al_{Y_{ijk}}\(-\f{\tl{d}^p_{mn}}{2}\al_{Y_{mnp}}+2c_r{\al_a}\)+2C_a(i) b_a \al^2_a\]2\pi^2\al\right\}\nn\\
 &+&2\pi \al M_0^2\[\f{d_{jk}^i}{2}\al_{Y_{ijk}}\(-1
+\f{1}{2\pi}\[\f{\tl{d}^p_{mn}}{2}\al_{Y_{mnp}}-2c_r^i {\al_i}\]2\pi^2\al \) +2C_a(i) \(\al_a-\f{1}{2\pi}b_a\al_a^2 2\pi^2\al\)\]\nn\\
&-&2\pi^3\al^2M_0^2\[\f{d_{jk}^i}{2}\f{1}{2\pi}\(\f{\tl{d}^p_{mn}}{2}\al_{Y_{mnp}}-2c_r^i{\al_i}\) \al_{Y_{ijk}}- 2C_a(i)\f{b_a}{2\pi}\al_a^2\]\nn\\
&=&0.
\eeqa
\normalsize
The subleading terms within $\pa^2 Z_i/\pa T^2$ are crucial for the exact cancelation of anomaly mediation and RGE effects.
\eit
So the numerical results of $'mirage'$ unification can be proved rigourously with our analytical expressions.

 \subsection{\label{sec-5}Deflection in Mirage Mediation From The Kahler Potential}
 It is well known that AMSB is bothered by tachyonic slepton problems. Such a problem in AMSB can be solved by the deflection of RGE trajectory 
with the introduction of messenger sector.  There are two possible ways to deflect the AMSB trajectory with the presence of messengers, 
either by pseudo-moduli field\cite{dAMSB} or
holomorphic terms (for messengers) in the Kahler potential\cite{Nelson:2002sa}.
Mirage mediation is a typical mixed modulus-anomaly mediation scenario. So the messenger sector, which can give additional gauge or Yukawa mediation contributions, can also be added in the Kahler potential.

 The Kahler potential involving the vector-like messengers $\bar{P}_i, P_i$ contain the ordinary kinetic terms as well as new holomorphic terms
\beqa\label{combine}
K\supseteq  \phi^\da\phi\[ Z_{P_i,\bar{P}_i}(T^\da,T)\( P_i^\da P_i+\bar{P}_i^\da \bar{P}_i\)+\(\tl{Z}_{P_i,\bar{P}_i}(T^\da,T)c_P \bar{P}_i P_i+h.c.\)\]~,
\eeqa
with
\beqa
Z_{P_i,\bar{P}_i}(T^\da,T)=\f{1}{(T+T^\da)^{{n}_{P}}}~,~~~\tl{Z}_{P_i,\bar{P}_i}(T^\da,T)=\f{1}{(T+T^\da)^{\tl{n}_{P}}}.
\eeqa
After normalizing and rescaling each superfield with the compensator field $\Phi \ra \phi \Phi$ and substituting the F-term VEVs of the compensator field 
$\phi=1+F_\phi\theta^2$, the relevant Kahler potential reduces to
\beqa
W=\int d^4 \theta \f{\phi^\da}{\phi}\f{1}{(T+T^\da)^{\tl{n}_P-n_P}} \left( c_P  R \bar{P}{P}\right),
\eeqa
For simply, we define $\tl{n}_P-n_P\equiv a_P$. Especially, $a_P=n_P$ for $\tl{n}_P=0$.

 The SUSY breaking effects can be taken into account by introducing a spurion superfields $R$ with
with the spurion VEV as
\beqa
R\equiv M_R+\theta^2 F_R=\f{1}{(2T)^{a_P}}\(F_\phi-\f{a_P}{2T}F_T\)+\theta^2 \[{a_P(a_P+1)}\f{ |F_T|^2}{4T^2}-|F_\phi|^2\].
\eeqa
with the value of the deflection parameter
\beqa
d\equiv\f{F_R}{M_R F_\phi}-1~,
\eeqa
depending on the choice of $a_P$ and $\al$ which gives $d=-2$ for $a_P=0$. 
We can see that adding messenger sector in the Kahler potential within mirage mediation will display a new feature in contrast to the AMSB case which always 
predicts $d=-2$.

The appearance of spurion messenger threshold will affect the AMSB RGE trajectory after integrating out the heavy messenger modes. The soft SUSY breaking parameters 
can be obtained by substituting $'d'$ into the general formula given in the appendix. Note that we can derive the final results directly with its low energy 
analytical expressions. Besides, we can also add messenger-matter mixing to induce new Yukawa couplings between the messengers and the MSSM fields. 
In this case, new Yukawa mediated contributions will also contribute to the low energy soft SUSY parameters (See Ref.\cite{fei2} for an example in AMSB).
 \subsection{\label{sec-5}Deflected Mirage Mediation With Messenger-Matter Interactions}
 In ordinary deflected mirage mediation SUSY breaking scenarios, additional messengers are introduced merely to amend the gauge beta functions which 
will subsequently feed into the low energy soft SUSY breaking parameters. In general, it is possible that the messengers will share some new Yukawa-type interactions 
with the visible (N)MSSM superfields, which subsequently will appear in the anomalous dimension of the superfields and contribute to 
the low energy soft SUSY breaking parameters. Such realizations have analogs in AMSB (see \cite{fei2}) and can be readily extended to include the 
modulus mediation contributions.

Similar to the deflected mirage mediation scenarios, the superpotential include possible pseudo-modulus superfields $X$,
the relevant nearly flat superpotential $W(X)$ to determine the deflection and a new part that includes messenger-matter interactions
 \beqa
 W_{mm}=\la_{\phi ij}X Q_iQ_j+y_{ijk}Q_i Q_j Q_k+W(X)~.
 \eeqa
with the Kahler potential
\beqa
K_m=Z_U\(T+T^\da,\f{\mu}{\sqrt{\phi^\da\phi}}\)\f{1}{(T+T^\da)^{n_{Q_i}}} Q_i^\da Q_i~,
\eeqa
Here $'\phi'$ denotes the compensator field with Weyl weight 1. The indices $'i,j'$ run over all MSSM and messenger fields 
and the subscripts $'U,D'$ denote the case upon and below the messenger threshold, respectively.

 After integrating out the heavy messenger fields, the visible sector superfields $Q_a$ will receive wavefunction normalization 
\beqa
{\cal L}=\int d^4\theta  Q_a^\da Z_D^{ab}(T+T^\da,\f{\mu}{\sqrt{\phi^\da\phi}},\sqrt{\f{X^\da X}{\phi^\da \phi}}) Q_b+\int d^2\theta y_{abc}Q^aQ^bQ^c~,
\eeqa
 which can give additional contributions to soft supersymmetry breaking parameters. Here the analytic continuing threshold superfield $'X'$ will 
trigger SUSY breaking mainly from the anomaly induced SUSY breaking effects with the form 
$<X>=M+\theta^2 F_X$. So we have
\beqa
\tl{X}\equiv \f{X}{\phi}&=& \f{M+F_X\theta^2}{1+F_\phi\theta^2}\equiv M(1+d F_\phi \theta^2),
\eeqa
with the value of the deflection parameter $'d'$ determined by the form of superpotential $W(X)$. 

Integrating out the messengers, the messenger-matter interactions will cause the discontinuity of the anomalous dimension upon and below the threshold. 
Such discontinuity will appear not only directly in the expressions for the trilinear couplings but also indirectly in the soft scalar masses. 
For example, the trilinear couplings at the messenger scale receive additional contributions
\beqa
\left.\Delta A_{ijk}\right|_{\mu=M}&=&\sum\limits_{a=i,j,k} \f{d}{2} F_\phi \f{\pa}{\pa\ln |X|} \left.\ln \[e^{-K_0/3}Z_a(\mu,X,T)\]\right|_{\mu=M}\nn\\
&=&\f{d}{2} F_\phi \sum\limits_{a=i,j,k}   \left.\Delta G_{i}\right|_{\mu=M}.
\eeqa
 We know that large trilinear couplings, especially $A_t$, is welcome in low energy phenomenological studies to reduce fine tuning and increase the Higgs mass. 
So the introduction of messenger-matter interactions can open new possibilities for mirage phenomenology.  

\section{\label{conclusion}Conclusions}
 We derive explicitly the soft SUSY breaking parameters at arbitrary low energy scale in the (deflected) mirage type mediation scenarios
with possible gauge or Yukawa mediation contributions. Based on the Wilsonian effective action after integrating out the messengers,
we obtain analytically the boundary value (at the GUT scale) dependencies of the effective wavefunctions and gauge kinetic terms.
Note that the messenger scale dependencies of the effective wavefunctions and gauge kinetic terms had already been discussed in GMSB. 
The RGE boundary value dependencies, which is a special feature in (deflected) mirage type mediation,
 is the key new ingredients in this study.
The appearance of $'mirage'$ unification scale in mirage mediation is proved rigorously with our analytical results.
We also discuss briefly the new features in deflected mirage mediation scenario in the case the deflection comes purely from the Kahler potential 
and the case with messenger-matter interactions.

We should note that our approach is in principle different from that of Ref.\cite{0901.0052} in which the soft SUSY breaking parameters are obtained
by numerical RGE evolution, matching and threshold corrections. For example, mixed gauge-modulus mediation contributions,
which will not appear in previous approach, will be necessarily present for the soft scalar masses in our approach.
\section*{Acknowledgement}
This work was supported by the Natural Science Foundation of China under grant numbers 11675147,11775012; by the Innovation Talent project of Henan Province under grant number 15HASTIT017.
\section*{Appendix A: Coefficients In Wavefunction Expansion}
We can construct the RGE invariants
\beqa
\f{d}{dt} \ln Z_i=\sum\limits_{l=y_t,y_b,y_\tau} A_l \f{d \ln y_{l}}{dt} +\sum\limits_{l=g_3,g_2,g_1} B_l \f{d \ln g_{l}}{dt}~,
\eeqa
by solving the equation in the basis of $(y_t^2,y_b^2,y_\tau^2,g_3^2,g_2^2,g_1^2)$
\beqa
\left(\bea{cccccc}
6&1&0&0&0&0\\1&6&3&0&0&0\\0&1&4&0&0&0\\-\f{16}{3}&-\f{16}{3}&0&b_3&0&0\\-3&-3&-3&0&b_2&0\\-\f{13}{15}&-\f{7}{15}&-\f{9}{5}&0&0&b_1
\eea\right)\left(\bea{c}
A_t\\A_b\\A_\tau\\B_3\\B_{2}\\B_{1}
\eea\right)=\left(\bea{c}
-2c_1\\-2c_2\\-2c_3\\-2d_1\\-2d_2\\-2d_3
\eea\right)~£¬
\eeqa
with $c_1,c_2,c_3,d_1,d_2,d_3$ the relevant coefficients of $(y_t^2,y_b^2,y_\tau^2,g_3^2,g_2^2,g_1^2)$ within the anomalous dimension.
So from
\beqa
\f{d}{dt}\[ Z_i(\mu)\prod\limits_{l=y_t,y_b,y_\tau} [y_l(\mu)]^{-A_l}\prod\limits_{k=1,2,3}[g_k(\mu)]^{-B_k}\]=0~,
\eeqa
we have
\beqa
Z_i(\mu)=Z_i(\Lambda)\prod\limits_{l=y_t,y_b,y_\tau} \(\f{y_l(\mu)}{y_l(\Lambda)}\)^{A_l}\prod\limits_{k=1,2,3}\(\f{g_k(\mu)}{g_k(\Lambda)}\)^{B_k}
\eeqa

The general expressions of wavefunction at ordinary scale $\mu$ below the messenger scale $M$ are given as
\beqa
Z_i(\mu)&=&Z_i(\Lambda)\prod\limits_{l=y_t,y_b,y_\tau} \(\f{y_l(M)}{y_l(\Lambda)}\)^{A_l}\prod\limits_{k=1,2,3}\(\f{g_k(M)}{g_k(\Lambda)}\)^{B_k}\prod\limits_{k=y_U} \(\f{y_k(M)}{y_k(\Lambda)}\)^{C_k}\nn\\
&&~~~~~~~~\prod\limits_{l=y_t,y_b,y_\tau} \(\f{y_l(\mu)}{y_l(M)}\)^{A_l^\pr}\prod\limits_{k=1,2,3}\(\f{g_k(\mu)}{g_k(M)}\)^{B_k^\pr}~,
\eeqa
with $y_U$ the interactions involving the messengers which will be integrated below the messenger scale.
The coefficients are listed in Table.\ref{NF=0} and Table.\ref{NF=1,2,4,5}.
\begin{table}[htbp]
\caption{ Relevant coefficients in wavefunction expansion with $N_F=0$ messengers.}
\begin{center}
\begin{tabular}{|c|c|c|c|c|c|c|}
\hline
&$A_1(y_t)$&$A_2(y_b)$&$A_3(y_\tau)$&$B_3(g_3)$&$B_2(g_2)$&$B_1(g_1)$\\
\hline
$Q_3$&$-\f{17}{61}$&-$\f{20}{61}$&$\f{5}{61}$&-$\f{128}{183}$&$\f{87}{61}$&-$\f{5}{183}$\\
\hline
$U_3$&$-\f{42}{61}$&~$\f{8}{61}$&-$\f{2}{61}$&-$\f{48}{61}$&$-\f{108}{61}$&~$\f{48}{671}$\\
\hline
$D_3$&$~\f{8}{61}$&-$\f{48}{61}$&$\f{12}{61}$&-$\f{112}{183}$&$-\f{84}{61}$&~$\f{112}{2013}$\\
\hline
$L_3$&$-\f{3}{61}$&~$\f{18}{61}$&$-\f{35}{61}$&-$\f{80}{183}$&$~\f{123}{61}$&-$\f{103}{2013}$\\
\hline
$E_3$&$-\f{6}{61}$&~$\f{36}{61}$&$-\f{70}{61}$&-$\f{160}{183}$&$-\f{120}{61}$&~$\f{160}{2013}$\\
\hline
$H_u$&$-\f{63}{61}$&~$\f{12}{61}$&$-\f{3}{61}$&~$\f{272}{183}$&$~\f{21}{61}$&-$\f{89}{2013}$\\
\hline
$H_d$&$~\f{9}{61}$&-$\f{54}{61}$&$-\f{17}{61}$&~$\f{80}{61}$&$-\f{3}{61}$&-$\f{19}{671}$\\
\hline
$Q_2$&~0&~0&~0&-$\f{16}{9}$&~3&~$\f{1}{99}$\\
\hline
$U_2$&~0&~0&~0&-$\f{16}{9}$&~0&~$\f{16}{99}$\\
\hline
$D_2$&~0&~0&~0&-$\f{16}{9}$&~0&~$\f{4}{99}$\\
\hline
$L_2$&~0&~0&~0&~0&~3&~$\f{1}{11}$\\
\hline
$E_2$&~0&~0&~0&~0&~0&~$\f{4}{11}$\\
\hline
\end{tabular}
\end{center}
\label{NF=0}
\end{table}

\begin{table}[htbp]
\caption{The coefficients with $N_F=0,1,2,4$ messengers without new Yukawa couplings involving the messengers-matter interactions.
 The coefficients for $y_t,y_b,y_\tau$, namely $A_1(y_t)$,$A_2(y_b)$,$A_3(y_\tau)$, are the same as the case $N_F=1$.}
\begin{center}
\begin{tabular}{|c|c|c|c|c|c|c|}
\hline
&$B_{g_3}(i)$&$B_{g_2}(i)$&$B_{g_1}(i)$\\
\hline
$Q_3$&$(-\f{128}{183},-\f{64}{61},-\f{128}{61},~\f{128}{61})$&$(\f{87}{61},~\f{87}{122},~\f{29}{61},~\f{87}{305})$&$(-\f{5}{183},-\f{55}{2318},-\f{55}{2623},-\f{55}{3233})$\\
\hline
$U_3$&$(-\f{48}{61},-\f{72}{61},-\f{144}{61},~\f{144}{61})$&$(-\f{108}{61},-\f{54}{61},-\f{36}{61},-\f{108}{305})$&$(~\f{48}{671},~\f{72}{1159},~\f{144}{2623},~\f{144}{3233})$\\
\hline
$D_3$&$(-\f{112}{183},-\f{56}{61},-\f{112}{61},~\f{112}{61})$&$(-\f{84}{61},-\f{42}{61},-\f{28}{61},-\f{84}{305})$&$(~\f{112}{2013},~\f{56}{1159},~\f{112}{2623},~\f{112}{3233})$\\
\hline
$L_3$&$(-\f{80}{183},-\f{40}{61},-\f{80}{61},~\f{80}{61})$&$(\f{123}{61},~\f{123}{122},~\f{41}{61},~\f{123}{305})$&$(-\f{103}{2013},-\f{103}{2318},-\f{103}{2623},-\f{103}{3233})$\\
\hline
$E_3$&$(-\f{160}{183},-\f{80}{61},-\f{160}{61},~\f{160}{61})$&$(-\f{120}{61},-\f{60}{61},-\f{40}{61},-\f{24}{61})$&$(\f{160}{2013},~\f{80}{1159},~\f{160}{2623},~\f{160}{3233})$\\
\hline
$H_u$&$(~\f{272}{183},~\f{136}{61},~\f{272}{61},-\f{272}{61})$&$(\f{21}{61},~\f{21}{122},~\f{7}{61},~\f{21}{305})$&$(-\f{89}{2013},-\f{89}{2318},-\f{89}{2623},-\f{89}{3233})$\\
\hline
$H_d$&$(~\f{80}{61},~\f{120}{61},~\f{240}{61},-\f{240}{61})$&$(-\f{3}{61},-\f{3}{122},-\f{1}{61},-\f{3}{305})$&$(-\f{19}{671},-\f{3}{122},-\f{57}{2623},-\f{57}{3233})$\\
\hline
$Q_2$&$(-\f{16}{9},-\f{8}{3},-\f{16}{3},~\f{16}{3})$&$(3,~\f{3}{2},~1,~\f{3}{5})$&$(\f{1}{99},~\f{1}{144},~\f{1}{129},~\f{1}{159})$\\
\hline
$U_2$&$(-\f{16}{9},-\f{8}{3},-\f{16}{3},~\f{16}{3})$&$(0,~0,~0,~0)$&$(\f{16}{99},~\f{8}{57},~\f{16}{129},~\f{16}{159})$\\
\hline
$D_2$&$(-\f{16}{9},-\f{8}{3},-\f{16}{3},~\f{16}{3})$&$(0,~0,~0,~0)$&$(\f{4}{99},~\f{2}{57},~\f{4}{129},~\f{4}{159})$\\
\hline
$L_2$&$(~0,~0,~0,~0)$&$(3,~\f{3}{2},~1,~\f{3}{5})$&$(\f{1}{11},~\f{3}{38},~\f{3}{43},~\f{3}{53})$\\
\hline
$E_2$&$(~0,~0,~0,~0)$&$(0,~0,~0,~0)$&$(\f{4}{11},~\f{6}{19},~\f{12}{43},~\f{12}{53})$\\
\hline
\end{tabular}
\end{center}
\label{NF=1,2,4,5}
\end{table}

\section*{Appendix B: Low Energy Spectrum in Deflected Mirage Mediation}
In order to show some essential features of our effective theory results, we list the predicted soft SUSY breaking parameters in deflected mirage 
mediation mechanism with $N_F$ messengers in ${\bf 5}\oplus \bar{\bf 5}$ representations of SU(5).

At energy $\mu$ below the messenger thresholds, we have
\bit
\item The gaugino masses:
\beqa
M_i(\mu)=l_a M_0\f{g_i^2(\mu)}{g_a^2(GUT)}+
\f{F_\phi}{16\pi^2}b_i g_i^2(\mu)-d \f{F_\phi}{16\pi^2}N_F g_i^2(\mu)~.
\eeqa
with
\beqa
(b_3~,b_2~,b_1)=(-3,~1,\f{33}{5}).
\eeqa

\item The trilinear couplings $A_t,A_b$ and $A_\tau$:

Note that at the messenger scale, the third contribution $\pa_X Z_i$ vanishes. 
The trilinear $A_t$ term is given at arbitrary low energy scale $\mu< M$ 
\beqa
&& A_t(\mu)-q_{y_t}M_0\\
&=&\f{M_0}{2\pi}\[6\f{q_{y_t}}{2}\al_{y_t}(\mu)+\f{q_{y_b}}{2}\al_{y_b}(\mu)-\f{16}{3}l_3\al_3(\mu)-3l_2\al_2(\mu)-\f{13}{15}l_1\al_1(\mu)\]\ln\(\f{M_{GUT}}{\mu}\)~\nn\\
&+&\f{F_\phi}{4\pi}\[6\al_{y_t}(\mu)+\al_{y_b}(\mu)-\f{16}{3}\al_3(\mu)-3\al_2(\mu)-\f{13}{15}\al_1(\mu) \]+\delta_G~,
\eeqa
Note that additional GMSB-type contributions are
\beqa
\delta_G
&=&d\f{F_\phi}{8\pi}\sum\limits_{k=1,2,3}\sum\limits_{F=Q_L^3,U_3,H_u}\[\f{}{}(B_k(F)-B_k^\pr(F)) (b_k+N_F) \al_k(M)+B_k^\pr(F) N_F \al_k(\mu)\]~\nn\\
&+&d\f{F_\phi}{8\pi}\sum\limits_{y_l=y_t,y_b,y_\tau}\sum\limits_{k=1,2,3}{A_l^\pr}\f{1}{4\pi^2} N_F\tl{c}_r(y_l)\al^2_r(\mu)\ln\(\f{M}{\mu}\) ~\nn\\
&=&d\f{F_\phi}{8\pi}\[-2\f{1}{8\pi^2} N_F\( \f{16}{3}\al^2_3(\mu)+3\al^2_2(\mu)+\f{13}{15}\al^2_1(\mu)\)\ln\(\f{M}{\mu}\)\]~,
\eeqa
with $2\tl{c}_r(y_l)$ the coefficients of $g_r^2$ within $-16\pi^2 \beta_{y_l}$ and
\beqa
\sum\limits_{F=Q_L^3,U_3,H_u}B_k(F)=\sum\limits_{F=Q_L^3,U_3,H_u}B_k^\pr(F)=0.
\eeqa
The trilinear $A_b$ term is
\beqa
&& A_b(\mu)-q_{y_b}M_0\\
&=&\f{M_0}{2\pi}\[\f{q_{y_t}}{2}\al_{y_t}(\mu)+6\f{q_{y_b}}{2}\al_{y_b}(\mu)+\f{q_{y_\tau}}{2}\al_{y_\tau}(\mu)-\f{16}{3}l_3\al_3(\mu)-3l_2\al_2(\mu)-\f{7}{15}l_1\al_1(\mu)\]\ln\(\f{M_{GUT}}{\mu}\)~\nn\\
&+&\f{F_\phi}{4\pi}\[\al_{y_t}(\mu)+6\al_{y_b}(\mu)+\al_{y_\tau}(\mu)-\f{16}{3}\al_3(\mu)-3\al_2(\mu)-\f{7}{15}\al_1(\mu) \]~\nn\\
&+&d\f{F_\phi}{8\pi}\[-2\f{1}{8\pi^2} N_F\( \f{16}{3}\al^2_3(\mu)+3\al^2_2(\mu)+\f{7}{15}\al^2_1(\mu)\)\ln\(\f{M}{\mu}\)\]~.
\eeqa
The trilinear $A_\tau$ term is
\beqa
&& A_\tau(\mu)-q_{y_\tau}M_0\\
&=&\f{M_0}{2\pi}\[3\f{q_{y_b}}{2}\al_{y_b}(\mu)+4\f{q_{y_\tau}}{2}\al_{y_\tau}(\mu)-3l_2\al_2(\mu)-\f{9}{5}l_1\al_1(\mu)\]\ln\(\f{M_{GUT}}{\mu}\)~\nn\\
&+&\f{F_\phi}{4\pi}\[3\al_{y_b}(\mu)+4\al_{y_\tau}(\mu)-3\al_2(\mu)-\f{9}{5}\al_1(\mu) \]~\nn\\
&+&d\f{F_\phi}{8\pi}\[-2\f{1}{8\pi^2} N_F\( 3\al^2_2(\mu)+\f{9}{5}\al^2_1(\mu)\)\ln\(\f{M}{\mu}\)\]~.
\eeqa
\item The soft SUSY breaking scalar masses are parameterized by several terms:
\beqa
&& -m_0^2=-(1-n_i)M_0^2+\delta_I+\delta_{II}+\delta_{III}+\delta_{IV}+\delta_{V}.~
\eeqa
The anomalous dimension of $Z_i$ is supposed to take the form
\beqa
G^i\equiv \f{d \ln Z_{i}}{d\ln\mu}=-\f{1}{2\pi }\(\f{1}{2}d_{kl}^i\al_{\la_{ikl}} -2C_a(i)\al_a\).
\eeqa
with $\al_{\la_{ikl}}=\la^2_{ikl}/4\pi$ and $\al_a=g_a^2/4\pi$.
\bit
\item Pure modulus mediation contributions
\beqa
\delta_I
&=&\f{M_0^2}{2\pi}\ln\(\f{M_{GUT}}{\mu}\)\left\{ \f{d_{jk}^i}{2}\(  q_{Y_{ijk}}^2+q_{Y_{ijk}}\)\al_{Y_{ijk}}(\mu)- 2C_a(i)\({l_a}+{l_a^2}\)\al_a\right.\nn\\
&+&\left.\f{1}{2\pi}\[\f{d_{jk}^i}{2}\al_{Y_{ijk}}(\mu)\(-\f{\tl{d}^p_{mn}}{2}q_{Y_{mnp}}\al_{Y_{mnp}}+2c_r l_a{\al_a}\)+2C_a(i) b_a \al^2_a\]
\ln\(\f{GUT}{\mu}\) \right\}\nn\\
\eeqa

\item Pure anomaly mediation contributions
\beqa
\delta_{II}&=&\f{F_\phi^2}{4}\f{\pa^2}{\pa (\ln \mu)^2}\ln \[e^{-K_0/3}Z_i\]\\
&=&-\f{F_\phi^2}{8\pi}\f{\pa}{\pa (\ln \mu)}\[\f{1}{2}d_{kl}^i\al_{\la_{ikl}} -2C_a(i)\al_a\]~,\nn\\
&=&-\f{F_\phi^2}{8\pi} \[\f{1}{2}d_{kl}^i \al_{\la_{ikl}}2 G^-_{\la_{ikl}} -2C_a(i)\f{1}{2\pi}b_a\al^2_a\].
\eeqa
with the beta function for Yukawa coupling $\la_{ikl}$ being
\beqa
\f{d \ln  \la_{ikl}}{d \ln\mu}=G_{\la_{ikl}}=\f{1}{4\pi}\[ \f{1}{2}{d^p_{mn}}\al_{\la_{mnp}}-2c_r \al_r \]~.
\eeqa
\item Pure gauge mediation contributions

As no new interactions involving the messengers are present, we have
\beqa
\f{\pa}{\pa \ln M}\ln \[e^{-K_0/3}Z_i\]&=&\f{1}{4\pi}\sum\limits_{g_k}\[(B_k-B_k^\pr) (b_k+N_F) \al_k(M)+B_k^\pr N_F \al_k(\mu)\]\nn\\
&+&\sum\limits_{Y_l} {A_l^\pr}\f{\tl{c}_r}{4\pi^2}  {\Delta b_r} \al_r^2(\mu)\ln\(\f{M}{\mu}\)~,
\eeqa
So
\beqa
\delta_{III}&=&d^2\f{F_\phi^2}{4}\f{\pa^2}{\pa (\ln M)^2}\ln \[e^{-K_0/3}Z_i\]\\
&=&d^2\f{F_\phi^2}{32\pi^2}\left\{\sum\limits_{g_k}\[(B_k-B_k^\pr) (b_k+N_F)^2 \al^2_k(M)
 +B_k^\pr N_F^2 \al^2(\mu)\]+\sum\limits_{Y_l} {A_l^\pr} \f{\tl{c}_r}{4\pi^2} {\Delta b_r}  \al_r^2(\mu)\right\}~.\nn
\eeqa
Here
\beqa
\f{\pa}{\pa \ln M}\al_k(M)&=&\f{b_k^+}{2\pi}\al_k(M)~,\nn\\
\f{\pa}{\pa \ln M}\al_k(\mu, M)&=&\f{b_k^+-b_k^-}{2\pi}\al_k(\mu,M)\equiv \f{\Delta b_k }{2\pi}\al_k(\mu,M)~,
\eeqa

\item  The gauge-anomaly interference term
\beqa
\delta_{IV}&=&-\f{d F^2_\phi}{2}\f{\pa^2}{\pa\ln M\pa\ln\mu}\ln \[e^{-K_0/3}Z_i(\mu,X,T)\]~,\nn\\
&=&-\f{d F_\phi^2}{2}\f{\pa}{\pa\ln\mu}\left\{\f{1}{4\pi}\sum\limits_{g_k}\[(B_k-B_k^\pr) (b_k+N_F) \al_k(M)+B_k^\pr N_F \al_k(\mu)\]\right.\nn\\
&&~~~~~~~~~~~~~~~+\left.\sum\limits_{Y_l} {A_l^\pr}\f{\tl{c}_r}{4\pi^2}  {\Delta b_r} \al_r^2(\mu)\ln\(\f{M}{\mu}\)\right\},~\nn\\
&=&-\f{d F_\phi^2}{16\pi^2} B_k^\pr b_k N_F \al^2_k(\mu)-d F_\phi^2\f{\tl{c}_r}{8\pi^2}{A_l^\pr}{\Delta b_r} \al_r^2(\mu).
\eeqa

\item  The modulus-anomaly and modulus-gauge interference terms are given as
\beqa
\delta_{V}&=&-\f{F_T F_\phi}{2}\f{\pa^2}{\pa T\pa\ln\mu}\ln \[e^{-K_0/3}Z_i\]+\f{d F_T F_\phi}{2}\f{\pa^2}{\pa T\pa\ln|X|}\ln \[e^{-K_0/3}Z_i\]~,\nn\\
&=&\f{F_T F_\phi}{4\pi}\f{\pa}{\pa T}\[\f{1}{2}d_{kl}^i\al_{\la_{ikl}}-2C_a(i)\al_a\]\nn\\
&+&\f{d F_T F_\phi}{2}\f{\pa}{\pa T}\left\{\f{1}{4\pi}\sum\limits_{g_k}\[(B_k-B_k^\pr) (b_k+N_F) \al_k(M)+B_k^\pr N_F \al_k(\mu)\]\right.\nn\\
&&~~~~~~~~~~~~~~~~~+\left.\sum\limits_{Y_l} {A_l^\pr}\f{\tl{c}_r}{4\pi^2}  {\Delta b_r} \al_r^2(\mu)\ln\(\f{M}{\mu}\)~\right\}~,\nn\\
&=&\f{M_0 F_\phi}{2\pi}\[\f{d_{kl}^i}{2}\al_{\la_{ikl}} \(-{q_{y_{\la_{ikl}}}}
+\f{1}{2\pi}\[\f{d^p_{mn}}{2} {q_{y_{\la_{mnp}}}}\al_{\la_{mnp}}-2c_r {l_r} \al_r\]\ln\[\f{M_G}{\mu}\]\)\right.\nn\\
 &&~~~~~~~~~\left.+ 2C_a(i)\f{l_a}{T} \f{\al_a^2}{\al_a(GUT)} \]\nn\\
&-&\f{d F_\phi M_0}{4\pi}\sum\limits_{g_k}\[(B_k-B_k^\pr) (b_k+N_F) l_k \f{ \al^2_k(M)}{\al_k(GUT)}+B_k^\pr N_F l_k \f{\al_k^2(\mu)}{\al_k(GUT)}\]\nn\\
&&~~~~~~~~~~~~~~~~~-\left.\sum\limits_{Y_l}{A_l^\pr}\f{d M_0 F_\phi }{4\pi^2} \tl{c}_r {\Delta b_r}l_r \f{2\al_r^3(\mu)}{\al_r(GUT)}\ln\(\f{M}{\mu}\)~\right\}~,
\eeqa
with
\beqa
\f{\pa}{\pa T}\al_k(\mu)=-\f{l_k}{T}\f{\al_k(\mu)}{\al_k(GUT)}\al_k(\mu)~,
\eeqa
\eit

\eit

\end{document}